\documentclass[twocolumn,twocolumn]{IEEEtran}
\usepackage[T1]{fontenc}
\usepackage{color}
\usepackage{mathrsfs}
\usepackage{amsmath}
\usepackage{amsthm} 
\usepackage{amssymb}
\usepackage{graphicx}
\usepackage{cite}
\usepackage{makecell}
\usepackage{algorithm}
\usepackage{algorithmic}
\usepackage{fancyhdr}
\usepackage{diagbox}
\usepackage{bm}
\usepackage{lettrine}
\usepackage{booktabs}  
\usepackage{subcaption}
\usepackage{caption} 
\usepackage{multirow} 
\usepackage{xcolor}

\usepackage{geometry}
\usepackage[
unicode=true,
bookmarks=true,
bookmarksnumbered=true,
bookmarksopen=true,
bookmarksopenlevel=1,
breaklinks=false,
pdfborder={0 0 0},
pdfborderstyle={},
backref=false,
colorlinks=true,  
linkcolor=blue,   
urlcolor=blue,    
citecolor=blue,   
anchorcolor=blue  
]{hyperref}

\hypersetup{
	pdftitle={Your Title},
	pdfauthor={Your Name},
	pdfpagelayout=OneColumn, 
	pdfnewwindow=true, 
	pdfstartview=XYZ, 
	plainpages=false
}

\usepackage{cleveref}

\allowdisplaybreaks[4]
\IEEEoverridecommandlockouts

\begin{document}

\title{\textcolor{black}{
DRO-Based Computation Offloading and Trajectory Design for Low-Altitude Networks}}
\author{
\IEEEauthorblockN{Guanwang Jiang$^{\dagger}$, Ziye Jia$^{\dagger}$, Can Cui$^{\dagger}$, Lijun He$^{\ast}$, 
Qiuming Zhu$^{\dagger}$, and
Qihui Wu$^{\dagger}$
\\
 }
\IEEEauthorblockA{
$^{\dagger}$Key Laboratory of Dynamic Cognitive System of Electromagnetic Spectrum Space, Ministry of Industry and 
Information Technology, Nanjing University of Aeronautics and  Astronautics, Nanjing, Jiangsu, 211106, China\\
$^{\ast}$School of Information and Control Engineering,
China University of Mining and Technology, \\Xuzhou, Jiangsu, 221116, China\\
\{gwjiang, jiaziye, cuican020619, zhuqiuming, wuqihui\}@nuaa.edu.cn, lijunhe.xd@gmail.com
}
\thanks{This work was supported in part by National Natural Science Foundation of China under Grant 62301251, in part by the Natural Science Foundation on Frontier Leading Technology Basic Research Project of Jiangsu under Grant BK20222001, in part by the Aeronautical Science Foundation of China 2023Z071052007, and in part by the Young Elite Scientists Sponsorship Program by CAST 2023QNRC001. ( \textit{Corresponding author: Ziye Jia}).}
}
\maketitle
\pagestyle{empty} 

\thispagestyle{empty}
\begin{abstract}
The low-altitude networks (LANs) integrating 
unmanned aerial vehicles (UAVs) and 
high-altitude platforms (HAPs) 
have become a promising solution 
for the rising computation demands. 
However, the uncertain task sizes and 
high mobility of UAVs pose great challenges to 
guarantee the quality of service. 
To address these issues, 
we propose an LAN architecture 
where UAVs and HAPs collaboratively provide computation offloading for ground users. 
Moreover, the uncertainty sets are constructed 
to characterize the uncertain task size, 
and
a distributionally robust optimization problem is formulated to 
minimize the worst-case delay by 
jointly optimizing the offloading decisions 
and UAV trajectories. 
To solve the mixed-integer min-max optimization problem, 
we design the distributionally robust computation offloading and 
trajectories optimization algorithm. 
Specifically, 
the original problem is figured out
by iteratively solving the outer-layer and inner-layer problems.
The convex outer-layer problem 
with probability distributions 
is solved by the optimization toolkit.
As for the inner-layer mixed-integer problem, 
we employ the Benders decomposition. 
The decoupled master problem concerning
the binary offloading decisions is solved by the integer solver,
and  UAV trajectories in the sub-problem 
are optimized via the successive convex approximation. 
Simulation results show 
the proposed algorithm outperforms 
traditional optimization methods 
in balancing the worst-case delay and robustness.
\end{abstract}

\section{INTRODUCTION}
\lettrine[lines=2]{T}HE development of the sixth generation communication technology 
has spawned a large number of on-demand computation services,
putting forward urgent  requirements for sufficient infrastructure, 
especially in remote areas.
Equipped with computation resources, 
the low-altitude network (LAN), composed of various aircrafts, 
can provide services for  these demands 
in infrastructure-limited areas\cite{5}. 
As a main component of LAN,
unmanned aerial vehicles (UAVs)
can adapt to dynamic user distributions 
due to the high flexibility and 
easy deployment \cite{9778241,jia2025hierarchical}. 
Meanwhile, high-altitude platforms (HAPs)
are worth boosting for their stable wide-area coverage, 
supplementing the limited endurance and coverage of UAVs. 
Therefore, the cooperation of UAVs and HAPs is
expected to provide on-demand and large-coverage 
computation  offloading services for ground users (GUs)\cite{6}. 

There exist some recent works related with LAN.
For example, the authors in \cite{10234543} optimized 
the UAV trajectories and task offloading ratios 
in the multi-UAV mobile edge computing system 
using the deep reinforcement learning (RL).
Authors in \cite{10118722} leveraged  the multi-agent RL
to reduce  the  energy consumption
in the scenario where the UAVs and HAP provided cooperative task offloading.
\cite{10937911} investigated the problem of 
the task assignment and resource optimization 
in an integrated multi-UAV system, 
aiming to maximize the sum of detection, tracking, and communication performance.
In most existing works, 
task sizes are assumed to
be fixed or predictable.
However,  
due to the
users' behaviors
and environmental factors, 
the task sizes exhibit stochastic variations\cite{3,10901605}. 
Although the traditional optimization methods 
offer ideas to  deal with 
the uncertainty, 
they struggle to 
balance the optimality and robustness\cite{beyer2007robust}. 
In contrast, the distributionally robust optimization (DRO)
can achieve a better trade-off by uncertainty sets\cite{pan2017data}. 

In this paper, we focus on the
joint optimization of the computation offloading decisions 
and UAV trajectories in LAN 
where UAVs and one HAP cooperate to 
provide computation offloading services for GUs.
To characterize the uncertain task sizes, 
we construct uncertainty sets based on 
the $L_1$ norm metric.
Then, a DRO problem is formulated to 
minimize the worst-case delay,
which is in the form of mixed-integer min-max optimization 
and hard to solve.
To deal with the problem,
we design the distributionally robust computation offloading 
and trajectories optimization (DRCOTO) algorithm.
Specially, we solve the
outer-layer minimization problem
and
inner-layer maximization problem iteratively.
The convex outer-layer problem 
related with the probability distribution 
is solved by the optimization solver.
To handle the mixed-integer inner-layer problem  
concerning  decisions and trajectories,
we employ
the Benders decomposition (BD) integrated with the
successive convex approximation (SCA). 
Finally, simulations 
evaluate the superiority of the proposed DRCOTO algorithm.

The rest of this paper is arranged as follows.
Section \ref{SYSTEM MODEL AND PROBLEM FORMULATION} presents the system model and problem formulation. 
Section \ref{Algorithm Design} details the DRCOTO algorithm. 
Section \ref{SIMULATION RESULTS} provides simulation results. 
Finally, Section \ref{CONCLUSIONS} concludes the paper.

\section{SYSTEM MODEL AND PROBLEM FORMULATION}\label{SYSTEM MODEL AND PROBLEM FORMULATION}
\begin{figure}[t]
	\centering
	\includegraphics[width=0.55\linewidth]{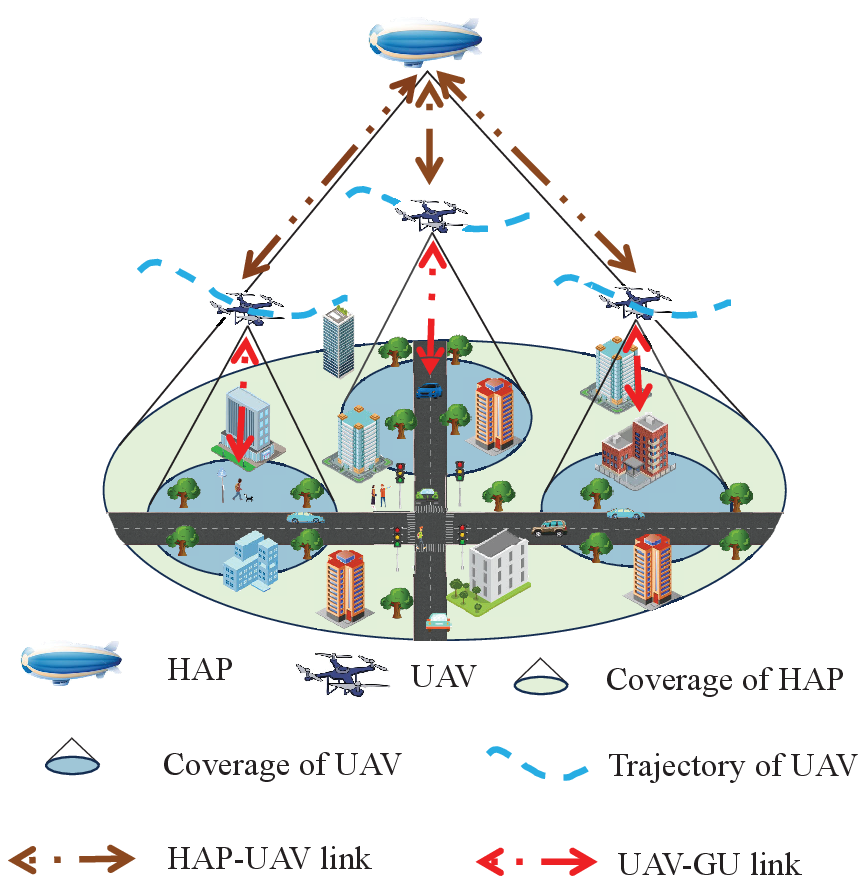}
	\caption{Scenario of LAN.}
	\label{fig:Scenario}
\end{figure}

As shown in Fig. \ref{fig:Scenario}, we consider an LAN
composed of GUs, cruising UAVs, and one hovering HAP, denoted as $i\in\mathcal{I}=\{1,2,...,I\}$, 
$j\in\mathcal{J}=\{1,2,...,J\}$, and  $H$, respectively. 
Both UAVs and the HAP carry edge servers to 
provide computing services for GUs.
The considered time period $\mathcal{T}$ is divided into 
$N$ time slots, indexed by 
$n \in \mathcal{N}=\{1,2,...,N\}$,
each with a duration of $\tau$. 
The slot length  $\tau$ is small enough so that 
the flying distance of UAV during each slot is negligible, 
and the channel gain is regarded as constant in each slot.  
The three-dimensional Cartesian coordinate 
is employed to describe the locations of GU $i$, UAV $j$ and HAP $H$ 
in time slot $n$, 
denoted as $q^{g}_{i}=(q^{gx}_{i}, q^{gy}_{i}, 0)$,  
$q^{u}_{j,n}=(q^{ux}_{j,n}, q^{uy}_{j,n}, q^{uz})$ 
and  $q^{h}=(q^{hx}, q^{hy}, q^{hz})$, respectively. 
Note that UAVs maintain flying 
with the fixed height $q^{uz}$, 
and the position of the HAP keeps unchanging.
\vspace{-0.3cm}
\subsection{Uncertainty Set for Task Size}
Each GU has a computation-intensive task $\phi_{i} =\{s_{i}, c_{i}\}$ 
to be processed 
over the entire time period $\mathcal{T}$.
$s_{i}$ represents the total task size measured in bits, 
and $c_{i}$ denotes the required number of CPU cycles per bit. 
To facilitate processing, each task $\phi_{i}$ is evenly divided into 
$N$ parts, with one part processed in each slot.
Hence, the data size of task $\phi_{i}$ to be computed in time
slot $n$ is $s_{i,n}=s_{i}/N$. 

In most scenarios, data size $s_{i}$ of task $\phi_{i}$ is uncertain,
and the probability distribution is unspecified. 
To model the uncertainty, 
each task size $s_i$ is supposed to 
have the same sample space $\Omega$,
which contains $K$ possible discrete values of the task
volume, i.e., $\Omega = \{s^k| \forall k=1,2,...,K\}\label{uncertainty set}$.
Meanwhile,  $s_i$
follows its respective probability distribution $\mathbb{P}_i=\{p_{i,k}| \forall k=1,2,...,K\}$,
where $p_{i,k}$ denotes the probability of sample $s^k$ in the distribution of $s_i$.
Besides, 
the reference distribution of $s_i$ is $\mathbb{P}^0_i=\{p_{i,k}^0| \forall k=1,2,...,K\}$, 
denoted as
\begin{equation}
	p_{i,k}^0=\frac{\sum^Q_{q=1}\delta^k(s_i)}{Q},\forall k=1,2,...,K.
\end{equation}
$Q$ is the number of the historical data and
$\sum^Q_{q=1}\delta^k(s_i)$ is 
the total  number of historical data size which falls in  interval 
$[d^k, d^{k+1})$.
In detail, 
if $d^k\leq  s_i  <d^{k+1}$, $\delta^k(s_i)$$=$$1$, 
and  otherwise $\delta^k(s_i)$$=$$0$. 
Moreover, 
we use the $L_1$ norm metric to 
quantify the distance between $\mathbb{P}^0_i$ and $\mathbb{P}_i$,
defined as
\begin{equation}
	d_{L_1}(\mathbb{P}^0_i,\mathbb{P}_i)=||\mathbb{P}^0_i - \mathbb{P}_i||_1=\sum\limits_{k=1}^K|p_{i,k}-p_{i,k}^0|\label{L1}.
\end{equation}
Then, the uncertainty set is
$\mathscr{D}_i=\{\mathbb{P}_i|d(\mathbb{P}^0_i,\mathbb{P}_i)\leq \epsilon \}\label{Di}$,
where $\epsilon$ is the tolerance value.

\subsection{Computation Model}
In the LAN scenario, GUs can either process their computation task
locally or offload them to  the UAV. 
Due to  the limited resources, 
UAVs can compute the collected data or
relay them  to the HAP for the further processing.
The decision variable $x_{i,j,n} \in \{0,1\}$ is  introduced. 
If $s_{i,n}$ is collected  by UAV $j$, $x_{i,j,n}=1$ 
and otherwise $x_{i,j,n}=0$. 
 The delay and energy for GU $i$ computing $s_{i,n}$ locally are
\begin{equation}
	T^{gcp}_{i,n}=\frac{\left(1-\sum\limits_{j=1}^{J}x_{i,j,n}\right)s_{i,n}c_{i}}{f^{gcp}},
\end{equation}
and
\begin{align}
	E^{gcp}_{i,n}=(1-\sum^{J}_{j=1}x_{i,j,n})\eta^{g}s_{i,n}c_{i}(f^{gcp})^2,
\end{align}
respectively.
$f^{gcp}$ and $\eta^{g}$ are the CPU cycles per second 
and effective capacitance coefficient of GU, respectively.
Similarly,
binary variable $y_{i,j,n}\in \{0,1\}$ is defined, 
and $y_{i,j,n}=1$ if $s_{i,n}$ is computed by UAV $j$, or $y_{i,j,n}=0$.
The delay and energy for UAV $j$ to compute data $s_{i,n}$ are
\begin{equation}
	T^{ucp}_{i,j,n}=\frac{s_{i,n}y_{i,j,n} c_{i}}{f^{ucp}},
\end{equation}
and
\begin{align}
	E^{ucp}_{i,j,n}=\eta^{u}y_{i,j,n}s_{i,n}c_{i}(f^{ucp})^2,
\end{align}
respectively.
$f^{ucp}$ and $\eta^{u}$ are the CPU cycles per second 
and effective capacitance coefficient of UAVs, respectively\cite{11006480}.
Besides, $z_{i,j,n}\in \{0,1\}$ is the index binary, 
and $z_{i,j,n}=1$ represents $s_{i,n}$ is relayed from UAV $j$ to HAP $H$.
The delay and energy for $s_{i,n}$ computed on HAP $H$ are
\begin{equation}
	T^{hcp}_{i,j,n}=\frac{s_{i,n}z_{i,j,n} c_{i}}{f^{hcp}},
\end{equation}
and
\begin{align}
	E^{hcp}_{i,j,n}=\eta^{h}s_{i,n}c_{i}(f^{hcp})^{2}z_{i,j,n},
\end{align}
respectively.
$f^{hcp}$ and $\eta^{h}$ are  the CPU cycles per second 
and effective capacitance coefficient of the HAP, respectively.

\vspace{-0.1cm}
\subsection{ Communication Model}
\subsubsection{GU to UAV}
The probabilistic  line-of-sight (LoS) channel model 
is leveraged for  the mixture of LoS and non-line-of-sight (NLoS)
environments between GUs and UAVs\cite{8833519}.
The probability
of the LoS link is 
$P_{LoS}(\theta_{i,j,n})=\frac{1}{1+aexp(-b(\theta_{i,j,n}-a))}$,
where $\theta_{i,j,n} = \arcsin\left( q^{uz} / d^{ug}_{i,j,n} \right)$ is the angle between GU $i$ and UAV $j$ in time slot $n$.
$d^{ug}_{i,j,n}$ is the Euclidean distance 
between GU $i$ and UAV $j$ in slot $n$.
$a$ and $b$ are the environment related parameters\cite{10118722}.
The probability of NLoS is
$P_{NLoS}(\theta_{i,j,n})=1-P_{LoS}(\theta_{i,j,n})$,
and the channel gain between
GU $i$ and UAV $j$  in time slot $n$ is
$g^{ug}_{i,j,n}=P_{LoS}(\theta_{i,j,n})\beta_{0}(d_{i,j,n}^{ug})^{-\alpha}
	+P_{NLoS}(\theta_{i,j,n})\kappa\beta_{0}(d_{i,j,n}^{ug})^{-\alpha}$,
where $\beta_{0}$, $\alpha$, and $\kappa$  are the path loss parameters.
According to the Shannon theory, the data transmission rate from GU $i$ to UAV $j$ in time slot $n$ is
\begin{equation}
	R^{ug}_{i,j,n}=B^{ug}\log_2\left(1+\frac{p^{gtr} g^{ug}_{i,j,n}}{\sigma^2+I^{ug}}\right).
\end{equation}
$B^{ug}$ represents the channel bandwidth between GUs and UAVs, 
$p^{gtr}$ denotes the maximum transmission power of GUs,
$\sigma^2$ is the noise power, 
and $I^{ug}$ is the average interference from other GUs, 
sharing  the same sub-channel\cite{10007714}.
Therefore, the delay for sending data  $s_{i,n}$ from GU $i$ to UAV $j$ is
\begin{equation}
	T^{gutr}_{i,j,n}=\frac{s_{i,n}x_{i,j,n}}{R^{ug}_{i,j,n}},
\end{equation}
and the corresponding energy cost is 
\begin{align}
	E^{gutr}_{i,n}=p^{gtr}\sum^{J}_{j=1}T^{gutr}_{i,j,n}.
\end{align}
Thus, the total energy consumption of GU $i$ to transmit and compute data $s_{i,n}$ is
\begin{align}
	E^{g}_{i,n}&=E^{gutr}_{i,n}+E^{gcp}_{i,n}.
\end{align}	

\subsubsection{UAV to HAP}
The free space path loss is considered to 
characterize communication links between 
UAVs and the HAP, 
ignoring the fading effects caused by 
the reflection and diffraction\cite{6261045}.
Besides, we apply the orthogonal frequency division multiplexing technique
to UAV-HAP links, and the mutual interference  can be avoided
\cite{8473695}. 
The achievable data transmission rate \cite{7} from UAV $j$ to HAP $H$ in time slot $n$ is
\begin{equation}
	R^{uh}_{j,n}=B^{uh}\log_2\left(1+\frac{p^{utr} G^{uh}L_{l}L^{uh}_{j}(n)}{B^{uh}K_{B}T_{s}}\right),
\end{equation}
where $B^{uh}$ is the bandwidth between the UAVs
and HAP, $G^{uh}$ is the antenna gain, $p^{utr}$ is the
transmit power of UAVs, $L_{l}$ is the total path loss, $K_{B}$
is the Boltzmann constant, $T_{s}$ is the system noise
temperature, and $L^{uh}_{j}(n) = \left(c/(4\pi d^{uh}_{j,n}f^{uh}_{c})\right)^2$
is the free-space path loss. 
$c$ is the speed of light, 
$d^{uh}_{j,n}$ is the Euclidean distance 
between the HAP and UAV $j$ in slot $n$, 
and $f^{uh}_{c}$ is the central frequency.
Then, the delay and energy consumption for sending  data $s_{i,n}$ from UAV $j$ to HAP $H$ are
\begin{equation}
	T^{uhtr}_{i,j,n}=\frac{s_{i,n}z_{i,j,n}}{R^{uh}_{j,n}},
\end{equation}
and
\begin{align}
	E^{uhtr}_{i,j,n}=p^{utr}T^{uhtr}_{i,j,n},
\end{align}
respectively.
Hence, 
the total delay for computing task $s_{i,n}$ is
\begin{align}
	T_{i,n}=T^{gcp}_{i,n}+\sum\limits_{j=1}^{J}\left(T^{gutr}_{i,j,n}+T^{uhtr}_{i,j,n}
	+T^{ucp}_{i,j,n}+T^{hcp}_{i,j,n}\right).
\end{align}

\subsection{UAV Trajectory Model}\label{Mobility Analysis of UAVs}
The UAV trajectories  are constrained  within the operation area, i.e.,
\begin{equation}
	0<q^{ux}_{j,n}<X_{max}, \forall j\in \mathcal{J}, n\in \mathcal{N},
	\label{cons of qxjn0}
\end{equation}
and
\begin{equation}
	0<q^{uy}_{j,n}<Y_{max}, \forall j\in \mathcal{J}, n\in \mathcal{N},
	\label{cons of qyjn0}
\end{equation}
where  $X_{max}$ and $Y_{max}$ are the horizontal boundaries.
UAVs are assumed to cruise with a
constant flight  speed  $v^{uf}$,
and the maximum horizontal flight distance during  each time slot is limited by 
\begin{equation}
	||q^{u}_{j,n}-q^{u}_{j,n-1}|| \leq v^{uf} \tau, \forall j\in \mathcal{J}, n\in \mathcal{N}.
	\label{cons of v}
\end{equation}
Considering the aerial safety, 
the inter-distance between any two UAVs shall be no less than $D_{min}$, i.e.,
\begin{equation}
	||q^{u}_{j,n}-q^{u}_{j^{'},n}|| \geq D_{min}, \forall j, j^{'}\in \mathcal{J}, j \neq j^{'}, n\in \mathcal{N}.
	\label{cons of Dmax}
\end{equation}

We consider the rotary wing UAV propulsion power model \cite{8663615} and 
the flight power $p^{uf}$ of UAV is 
\begin{align}\notag
	p^{uf}&=P_{1}\left(1+\frac{3||v^{uf}||^{2}}{U^{2}_{tip}}\right)+\frac{1}{2}d_{0}\varsigma_{0}sA||v^{uf}||^{3}\\
	&+P_{2}\left(\sqrt{1+\frac{||v^{uf}||^{4}}{4v_{0}^{2}}}-\frac{||v^{uf}||^{2}}{2v_{0}^{2}}\right)^{\frac{1}{2}},
\end{align}
where $P_{1}$ is the power of the UAV blade, $U_{tip}$ represents
the blade tip speed, $d_{0}$ denotes the fuselage drag ratio, $\varsigma_{0}$ is the air density, 
$s$ is the rotor solidity,  $A$ is the rotor
area, $P_{2}$ is the
induced power during hovering, and
$v_{0}$ is the mean velocity of rotors. 
Moreover, the hovering power consumption of UAVs is
\begin{align}
	p^{uhov}=P_{1}+P_{2}.
\end{align}
The flight time for UAV $j$  during time slot $n$ is  calculated as
\begin{align}
	T^{fly}_{u,n}=\frac{||q^{u}_{j,n}-q^{u}_{j,n-1}||}{v^{uf}}.
\end{align}
The propulsion energy cost of UAV $j$ in time slot $n$ is 
\begin{align}
	E^{uf}_{j,n}=p^{uf} T^{fly}_{u,n}+p^{uhov}(\tau-T^{fly}_{u,n}).
\end{align}

The energy consumption $E_{j}^{u}$ of UAV $j$ 
mainly consists of three parts: the communication cost $E^{uhtr}_{i,j,n}$, 
computing cost $E^{ucp}_{i,j,n}$, and movement cost $E^{uf}_{j,n}$, i.e.,
\begin{align}
	E^{u}_{j}&=\sum\limits_{i=1}^{I}\sum\limits_{n=1}^{N}E^{uhtr}_{i,j,n}+\sum\limits_{i=1}^{I}\sum\limits_{n=1}^{N}E^{ucp}_{i,j,n}+\sum\limits_{n=1}^{N}E^{uf}_{j,n}.
\end{align}

\vspace{-0.5cm}
\subsection{Problem Formulation}\label{Problem Formulation}
The DRO problem $\mathbf{P0}$ is formulated 
on the basis of the constructed uncertainty sets, 
jointly optimizing the computation offloading strategies and UAV trajectories,
to minimize the expected maximum  total worst-case delay, i.e.,
\begin{align}
	\notag \mathbf{P0}\textrm{:}\;&\underset{\boldsymbol{x},\boldsymbol{y},\boldsymbol{z},\boldsymbol{q}}{\textrm{min}}\underset{\boldsymbol{p}}{\textrm{max}}\sum\limits_{i=1}^{I}\sum\limits_{n=1}^{N}\mathbb{E}_{\mathbb{P}_{i}}\left(T_{i,n}\right)\\
	\notag \textrm{s.t.}\;
	&(\ref{cons of qxjn0})-(\ref{cons of Dmax}),\\
	&\sum\limits_{j=1}^{J}x_{i,j,n}\leq 1, \forall i\in \mathcal{I},  n\in \mathcal{N}, \label{one i one xij}\\
	&\sum\limits_{i=1}^{I}y_{i,j,n}\leq N^{u}, \forall j\in \mathcal{J},  n\in \mathcal{N}, \label{cons of NU}\\
	&\sum\limits_{i=1}^{I}\sum\limits_{j=1}^{J}z_{i,j,n}\leq N^{h}, \forall n\in \mathcal{N}, \label{cons of NH}\\
	&y_{i,j,n} + z_{i,j,n} = x_{i,j,n} , \forall i\in \mathcal{I},j\in \mathcal{J}, n\in \mathcal{N}, \label{cons of x,  y, and z}\\
	&\mathbb{E}_{\mathbb{P}_{i}}\left(T_{i,n}\right)\leq \tau, \forall i\in \mathcal{I}, n\in \mathcal{N},\label{cons of Tin}\\
	&\sum\limits_{n=1}^{N}\mathbb{E}_{\mathbb{P}_{i}}\left(E^{g}_{i,n}\right) \leq E^{gmax},\forall i\in\mathcal{I},\label{cons of EGmax}\\
	&\mathbb{E}_{\mathbb{P}_{i}}\left(E^{u}_{j}\right) \leq E^{umax},\forall j\in\mathcal{J},\label{cons of EUmax}\\
	&\sum\limits_{i=1}^{I}\sum\limits_{j=1}^{J}\sum\limits_{n=1}^{N}\mathbb{E}_{\mathbb{P}_{i}}\left(E^{hcp}_{i,j,n}\right) \leq E^{hmax},\label{cons of EHmax}\\
	&\mathbb{P}_{i}\in{\mathscr{D}_{i}}, \forall i\in\mathcal{I}, n\in \mathcal{N},\label{cons of Pi}\\
	&x_{i,j,n}, y_{i,j,n}, z_{i,j,n}\in\{0,1\},\forall i\in\mathcal{I}, j\in\mathcal{J}, n\in \mathcal{N}, \label{cons of z}
\end{align}
where  
$\mathbb{E}_{\mathbb{P}_{i}}\left(T_{i,n}\right)$ represents the expectation  of $T_{i,n}$  under the  probability  distribution $\mathbb{P}_{i}$.
$\boldsymbol{x} = \{x_{i,j,n},\forall i\in\mathcal{I}, j\in \mathcal{J}, n\in \mathcal{N}\}$,
$\boldsymbol{y} = \{y_{i,j,n},\forall i\in\mathcal{I}, j\in \mathcal{J}, n\in \mathcal{N}\}$,
$\boldsymbol{z} = \{z_{i,j,n},\forall i\in\mathcal{I}, j\in \mathcal{J}, n\in \mathcal{N}\}$,
$\boldsymbol{q}= \{ q^{u}_{j,n},\forall j\in\mathcal{J}, n\in \mathcal{N}\}$,
and
$\boldsymbol{p}= \{\mathbb{P}_{i},\forall i\in\mathcal{I}\}$.
Constraint (\ref{one i one xij}) dictates that 
each task can be collected by at most one UAV during each time slot.
Constraints (\ref{cons of NU}) and (\ref{cons of NH}) 
impose the numbers of GUs served by the UAV and  HAP
cannot exceed  $N^{u}$ and $N^{h}$, respectively.
Constraint (\ref{cons of x,  y, and z}) indicates the data flow balancing on the UAV. 
Constraint (\ref{cons of Tin}) specifies that 
the delay for completing processing  data  $s_{i,n}$ must not surpass the duration of time slot $\tau$.
Constraints (\ref{cons of EGmax}), (\ref{cons of EUmax}) and (\ref{cons of EHmax}) 
confine the energy consumption of each GU, UAV, and HAP 
to stay within the maximum capacity limits of  
$E^{gmax}$, $E^{umax}$, and $E^{hmax}$, respectively. 
Constraint (\ref{cons of Pi}) implies that 
probability distribution $\mathbb{P}_i$ for 
task size $s_{i}$ belongs to uncertainty set ${\mathscr{D}_i}$. 

Note that $\mathbf{P0}$ is a mix-integer problem, 
concerning binary variables  $\boldsymbol{x}$, 
$\boldsymbol{y}$, and $\boldsymbol{z}$,
as well as continuous variables $\boldsymbol{p}$ and $\boldsymbol{q}$.
Since the time complexity is
exponential with the problem scale growing,
 it is intractable to solve
$\mathbf{P0}$ efficiently.

\vspace{-0.3cm}
\section{Algorithm Design}\label{Algorithm Design}
To address the great challenges in solving $\mathbf{P0}$, 
we design the DRCOTO algorithm based the BD and SCA.
First,
 we discretize the sample space
into $K$, denoted as $\Omega = \{s^{k}|\forall k = 1, 2, ...,K\}$, 
and problem $\mathbf{P0}$ is reformulated as 
\begin{align}
	\notag \mathbf{P1}\textrm{:}\;&\underset{\boldsymbol{x},\boldsymbol{y},\boldsymbol{z},\boldsymbol{q}}{\textrm{min}}\underset{\boldsymbol{p}}{\textrm{max}}\sum\limits_{i=1}^{I}\sum\limits_{n=1}^{N}\sum\limits_{k=1}^{K}p_{i,k}T_{i,n,k}\\
	\notag \textrm{s.t.}\;
	&(\ref{cons of qxjn0})-(\ref{cons of Dmax}), (\ref{one i one xij})-(\ref{cons of z}).
\end{align}
Based on the discretization, 
$s_{i,n}$ in $T_{i,n}$ can be replaced by $s^k$, 
and $T_{i,n}$ is transformed into $T_{i,n,k}$.
With the fixed $p_{i,k}$ of the inner-layer max problem, 
$\mathbf{P2}$ is given by
\begin{align}
	\notag \mathbf{P2}\textrm{:}\;&\underset{\boldsymbol{x},\boldsymbol{y},\boldsymbol{z}, \boldsymbol{q}}{\textrm{min}}\sum\limits_{i=1}^{I}\sum\limits_{n=1}^{N}\sum\limits_{k=1}^{K}p_{i,k}T_{i,n,k}\\
	\notag \textrm{s.t.}\;
	&(\ref{cons of qxjn0})-(\ref{cons of Dmax}), (\ref{one i one xij})-(\ref{cons of EHmax}), (\ref{cons of z}).
\end{align}
By solving problem $\mathbf{P2}$, 
$\boldsymbol{x}$, $\boldsymbol{y}$, $\boldsymbol{z}$, and $\boldsymbol{q}$ 
are obtained. 
With the pre-determined $\boldsymbol{x}$, $\boldsymbol{y}$, $\boldsymbol{z}$, 
and $\boldsymbol{q}$, 
$\mathbf{P1}$ is transformed to $\mathbf{P3}$, i.e.,
\begin{align}
	\notag \mathbf{P3}\textrm{:}\;&\underset{\boldsymbol{p}}{\textrm{max}}\sum\limits_{i=1}^{I}\sum\limits_{n=1}^{N}\sum\limits_{k=1}^{K}p_{i,k}T_{i,n,k}\\
	\notag \textrm{s.t.}\;
	&(\ref{cons of Tin})-(\ref{cons of Pi}).
\end{align}
Then, we handle $\mathbf{P1}$ by 
addressing $\mathbf{P2}$ and $\mathbf{P3}$ through alternating iterations.

To tackle the coupled continuous
and integer variables,
$\mathbf{P2}$
is decomposed into
a sub-problem $\mathbf{SP}$ and 
a master problem $\mathbf{MP}$ based on the BD.
Specifically,
given binary variables $\boldsymbol{x}$, $\boldsymbol{y}$, and $\boldsymbol{z}$ 
generated by $\mathbf{MP}$ at the
$\omega$-th iteration, $\mathbf{SP}$ is
\begin{align}
	\notag \mathbf{SP}\textrm{:}\;&\underset{\boldsymbol{q}}{\textrm{min}}\sum\limits_{i=1}^{I}\sum\limits_{n=1}^{N}\sum\limits_{k=1}^{K}p_{i,k}T_{i,n,k}\\
	\notag \textrm{s.t.}\;
	&(\ref{cons of qxjn0})-(\ref{cons of Dmax}), (\ref{cons of Tin})-(\ref{cons of EUmax}).
\end{align}
To deal with non-convex problem  $\mathbf{SP}$,  
we design the  algorithm based on the SCA to 
achieve a local optimal solution. 
Specifically, at the $m$-th iteration of the SCA, 
we implement the first-order Taylor series
expansions of $T_{i,n,k}$ and $E^{uf}_{j,n}$ to obtain the approximated functions
$\hat{T}^{m}_{i,n,k}$ and $\hat{E}^{muf}_{j,n}$, i.e.,
\begin{equation}
	\hat{T}^{m}_{i,n,k}={T}_{i,n,k}(\boldsymbol{{q}}^{m-1})+\nabla{T}_{i,n,k}(\boldsymbol{{q}}^{m-1})(\boldsymbol{{q}}^{m}-\boldsymbol{{q}}^{m-1}),
\end{equation}
and
\begin{equation}
	\hat{E}^{muf}_{j,n}={E}^{uf}_{j,n}(\boldsymbol{{q}}^{m-1})+\nabla{E}^{uf}_{j,n}(\boldsymbol{{q}}^{m-1})(\boldsymbol{{q}}^{m}-\boldsymbol{{q}}^{m-1}),
\end{equation}
respectively.
$\boldsymbol{{q}}^{m-1}$ is the solution obtained at the ($m-1$)-th iteration of the SCA. 
${T}_{i,n,k}(\boldsymbol{{q}}^{m-1})$ and ${E}^{uf}_{j,n}(\boldsymbol{{q}}^{m-1})$ 
are values of ${T}_{i,n,k}$ and ${E}^{uf}_{j,n}$ at the point $\boldsymbol{{q}}^{m-1}$, respectively.
Additionally, $\nabla f(x)$ represents the derivative of $f(x)$ at $x$.
Then, at the $m$-th iteration of the SCA, problem $\mathbf{SP}$ is transformed into
\begin{align}
	\notag \mathbf{SP'}\textrm{:}\;&\underset{\boldsymbol{q}}{\textrm{min}}\sum\limits_{i=1}^{I}\sum\limits_{n=1}^{N}\sum\limits_{k=1}^{K}p_{i,k}\hat{T}^{m}_{i,n,k}\\
	\notag \textrm{s.t.}\;
	&(\ref{cons of qxjn0})-(\ref{cons of Dmax}), (\ref{cons of Tin})-(\ref{cons of EUmax}).
\end{align}
The Lagrangian function of problem $\mathbf{SP'}$ at the $\omega$-th iteration of the BD is denoted as
\begin{align}
	\notag &L\left(\boldsymbol{{x}}^{\omega-1},\boldsymbol{{y}}^{\omega-1},\boldsymbol{{z}}^{\omega-1},\boldsymbol{q}\right)\\
	&=\sum\limits_{i=1}^{I}\sum\limits_{n=1}^{N}\sum\limits_{k=1}^{K}p_{i,k}T_{i,n,k}
	+(\boldsymbol{\lambda}^{\omega})^{T}(G\left(\boldsymbol{{x}}^{\omega-1},\boldsymbol{{y}}^{\omega-1},\boldsymbol{{z}}^{\omega-1},\boldsymbol{q}\right)),
\end{align}
where $\boldsymbol{\lambda}^{\omega}$ is the vector of dual factors, 
$(\boldsymbol{\lambda}^{\omega})^{T}$ is the transpose of $\boldsymbol{\lambda}^{\omega}$ and  $G\left(\boldsymbol{{x}}^{\omega-1},\boldsymbol{{y}}^{\omega-1},\boldsymbol{{z}}^{\omega-1},\boldsymbol{q}\right)$
is the constraint sets for problem $\mathbf{SP'}$,
defined as
\begin{gather}
	\begin{aligned}
		&G\left(\boldsymbol{{x}}^{\omega-1},\boldsymbol{{y}}^{\omega-1},\boldsymbol{{z}}^{\omega-1},\boldsymbol{q}\right)= \\
		&\left\{
		\begin{aligned}
			&-q^{ux}_{j,n}, 
			-q^{uy}_{j,n}, \forall j\in \mathcal{J}, n\in \mathcal{N},\\
			&q^{ux}_{j,n}- X_{max}, 
			q^{uy}_{j,n}- Y_{max}, \forall j\in \mathcal{J}, n\in \mathcal{N},\\
			&\|q^{u}_{j,n}-q_{j,n-1}\| - v^{uf} \tau, \forall j\in \mathcal{J}, n\in \mathcal{N},\\
			&D_{min} - \|q^{u}_{j,n}-q_{j^{'}}(n)\| , \forall j, j^{'}\in \mathcal{J}, j \neq j^{'}, n\in \mathcal{N},\\
			&\mathbb{E}_{\mathbb{P}_{i}}\left(T_{i,n,k}\right)- \tau, \forall i\in \mathcal{I}, n\in \mathcal{N},\\
			&\sum\limits_{n=1}^{N}\mathbb{E}_{\mathbb{P}_{i}}\left(E^{g}_{i,n}\right)- E^{gmax},\forall i\in\mathcal{I},\\
			&\mathbb{E}_{\mathbb{P}_{i}}\left(E^{U}_{j}\right) - E^{umax},\forall j\in\mathcal{J}.
		\end{aligned}
		\right\}
	\end{aligned}
\end{gather}
\begin{algorithm}[t]
	\caption{DRCOTO Algorithm}
	\label{DRCOTO Algorithm}
	\textbf{Initial:} $r=\omega=m=0$, 
	$D_{1}^{r}=D_{2}^{\omega}=D_{3}^{m}=\text{UB}=+\infty$, $\text{LB} = -\infty$, 
	$\boldsymbol{x}^{0}$, $\boldsymbol{y}^{0}$, $\boldsymbol{z}^{0}$,
	$\boldsymbol{p}^{0}$, 
	$\boldsymbol{q}^{0}$. 
	\begin{algorithmic}[1]\label{algorithm 1}
		\REPEAT
		\STATE $r=r+1$.\label{line 2}
		\STATE Substitute $\boldsymbol{{p}^{0}}$ into $\mathbf{P1}$ to get $\mathbf{P2}$.\label{line 3}
		\REPEAT
		\STATE $\omega=\omega+1$.
		\STATE Substitute $\boldsymbol{x}^{0}$, $\boldsymbol{y}^{0}$, and $\boldsymbol{z}^{0}$ into $\mathbf{P2}$ to get $\mathbf{SP}$.\label{line 6}
		\REPEAT \label{line 7}
		\STATE $m = m + 1$.\label{line 8}
		Perform the first-order Taylor expansion of 
		$T_{i,n,k}$ and $E^{uf}_{j,n}$ at point $\boldsymbol{q}^{0}$.
		\STATE Solve $\mathbf{SP'}$ to get $\boldsymbol{{q}}^{m}$ and the delay $D_{3}^{m}$.
		$\boldsymbol{q}^{0}=\boldsymbol{q}^{m}$.
		\UNTIL {$|D_{3}^{m-1}-D_{3}^{m}|$ $\leq \varrho$} or $m=m^{\max}$.\label{line 12}
		\STATE  UB = min\{UB, $D_{3}^{m}$\}. $\boldsymbol{{q}}^{\omega}=\boldsymbol{{q}}^{m}$. \label{line 14}
		\STATE Calculate the Benders cut according to (\ref{cut}) and then
		add it to master problem $\mathbf{MP}$.\label{line 15}
		\STATE Solve $\mathbf{MP}$ to get $\boldsymbol{{x}}^{\omega}$, $\boldsymbol{{y}}^{\omega}$, $\boldsymbol{z}^{\omega}$ and the delay $D_{2}^{\omega}$.\label{line 16}
		\STATE $\boldsymbol{x}^{0}=\boldsymbol{{x}}^{\omega}$, $\boldsymbol{y}^{0}=\boldsymbol{{y}}^{\omega}$, and $\boldsymbol{z}^{0}=\boldsymbol{{z}}^{\omega}$.
		\STATE LB = max\{LB, $D_{2}^{\omega}$\}.\label{line 18}
		\UNTIL UB $-$ LB $\leq$ $\delta$ or $\omega=\omega^{\max}$.
		\STATE Substitute $\boldsymbol{x}^{\omega}$, $\boldsymbol{y}^{\omega}$, $\boldsymbol{z}^{\omega}$, $\boldsymbol{q}^{\omega}$ into $\mathbf{P1}$ to get $\mathbf{P3}$.\label{line 21}
		\STATE Solve $\mathbf{P3}$ by the optimizer to get $\boldsymbol{{p}^{r}}$ and the delay $D_{3}^{r}$.\label{line 22}
		\STATE  $\boldsymbol{p}^{0}=\boldsymbol{{p}}^{r}$.\label{line 23}
		\UNTIL $|D_{3}^{r-1}-D_{3}^{r}| \leq \zeta$ or $r=r^{max}$.
	\end{algorithmic}
	\textbf{Output:} $\boldsymbol{x}^{\omega}$, $\boldsymbol{y}^{\omega}$, $\boldsymbol{z}^{\omega}$, $\boldsymbol{q}^{\omega}$, and $\boldsymbol{p}^{r}$.
\end{algorithm}
After solving $\mathbf{SP}$, we obtain the optimality cut at the $\omega$-th iteration, 
and  $\mathbf{MP}$ at the $\omega$-th iteration is
\begin{align}
	\notag \mathbf{MP}\textrm{:}\;&\underset{\boldsymbol{x},\boldsymbol{y},\boldsymbol{z}}{\textrm{min}}\xi \\
	\notag \textrm{s.t.}\;
	&(\ref{one i one xij})-(\ref{cons of EHmax}), (\ref{cons of z}),\\
	&L(\boldsymbol{x},\boldsymbol{y},\boldsymbol{z},\boldsymbol{{q}}^{\omega}) \leq \xi. \label{cut}
\end{align}
By adding Benders cuts as constraint (\ref{cut}), the search
space for the global optimal solution is gradually reduced. 
Then, we can solve problem $\mathbf{MP}$
using the optimization tool such as Gurobi.

The overall algorithm of the DRCOTO is summarized in Algorithm \ref{DRCOTO Algorithm}.
First, the decision variables
$\boldsymbol{x}^{0}$, $\boldsymbol{y}^{0}$, $\mathbf{z}^{0}$, $\boldsymbol{p}^{0}$, 
and $\boldsymbol{q}^{0}$  are  initialized as the arbitrary feasible solutions.
At the $r$-th iteration, 
$\boldsymbol{p}^{0}$ is fixed
in $\mathbf{P1}$ to 
get problem $\mathbf{P2}$ (line \ref{line 3}).
Then, we substitute 
$\boldsymbol{x}^{0}$, $\boldsymbol{y}^{0}$, and $\boldsymbol{z}^{0}$ 
into $\mathbf{P2}$ to get problem $\mathbf{SP}$ (line \ref{line 6}). 
To solve problem $\mathbf{SP}$, 
the SCA is leveraged (lines \ref{line 7}-\ref{line 12}). 
We  update the upper bound of problem $\mathbf{P2}$
with objective value $D_{3}^{m}$ of problem $\mathbf{SP'}$ (line \ref{line 14}).
Then, the Benders cut according to (\ref{cut}) is calculated and
added to master problem $\mathbf{MP}$ (line \ref{line 15}).
By solving problem $\mathbf{MP}$, 
the lower bound of problem $\mathbf{P2}$ is updated (lines \ref{line 16}-\ref{line 18}).
Repeat the steps \ref{line 8}-\ref{line 18} until the BD is convergent.
Besides, $\mathbf{P1}$ is transformed into $\mathbf{P3}$
by substituting variables $\boldsymbol{x}^{\omega}$, $\boldsymbol{y}^{\omega}$, $\boldsymbol{z}^{\omega}$, $\boldsymbol{q}^{\omega}$ 
and UAV trajectories into problem $\mathbf{P1}$ (line \ref{line 21}).
By solving problem $\mathbf{P3}$, 
the possible distribution of task sizes is updated (lines \ref{line 22}-\ref{line 23}).
Repeat the steps \ref{line 2}-\ref{line 23} until the DRCOTO converges.
Finally, we get the computation offloading related variables 
and UAV trajectories.
\begin{figure}[t]
	\centering
	\includegraphics[width=0.8\linewidth]{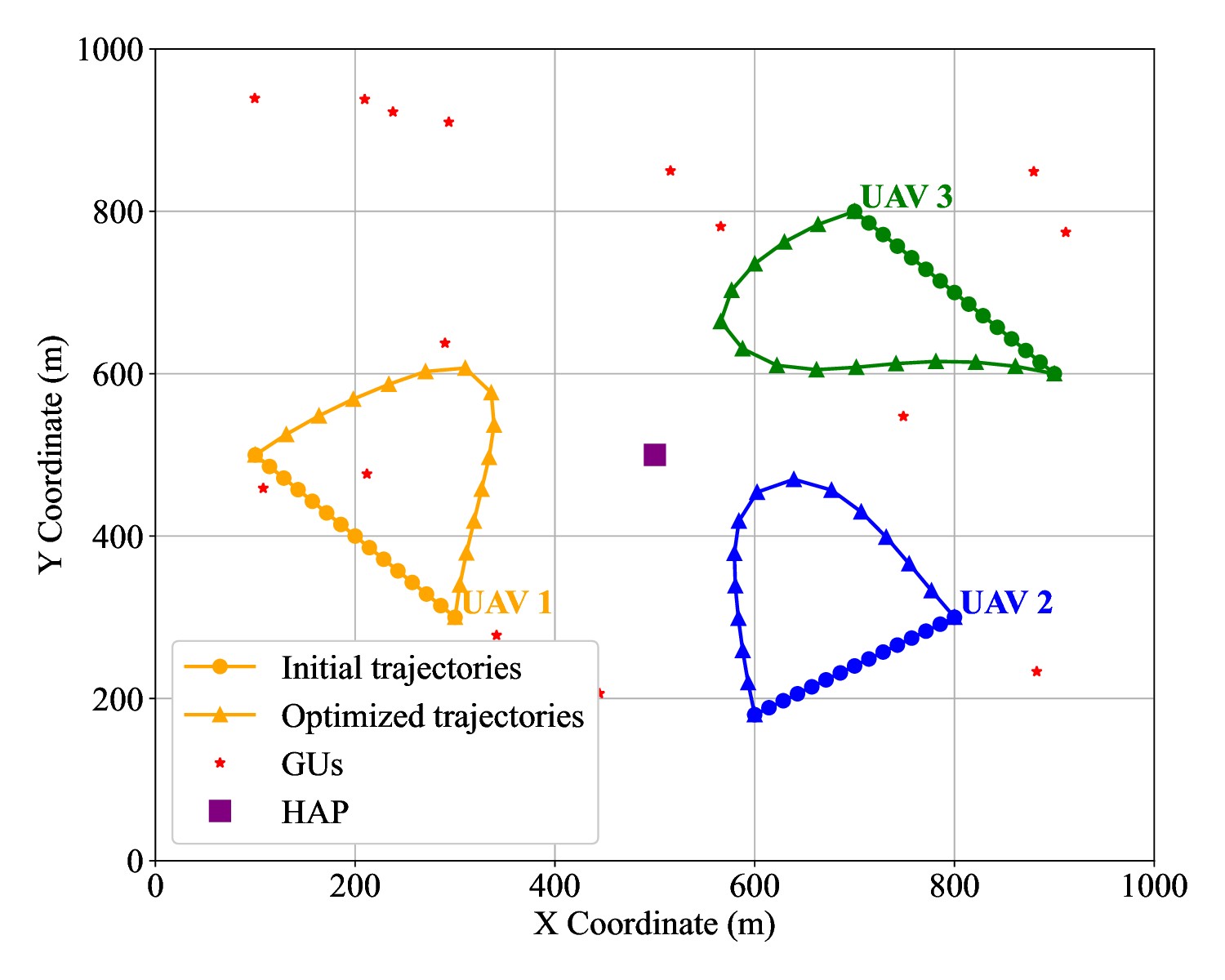}
	\caption{The  UAV trajectories and locations of GUs and HAP.}
	\label{fig:12}
\end{figure}

\vspace{-0.3cm}
\section{SIMULATION RESULTS}\label{SIMULATION RESULTS}
We consider an LAN scenario 
within a $X_{max}\times Y_{max}=1$km$\times$ $1$km ground area.
$I=15$ GUs are  randomly distributed.
$J=3$ UAVs and $1$ HAP collaboratively provide computation offloading service.
The sample space of the task size is $\Omega=\{0.2,0.5,1,1.5,2\}$Mbit. 
The radius of the uncertainty set is $\epsilon=0.3$. 
The key parameters are set as:
$I^{ug}=-90$dBm,
$N=15$,
$\tau=2$s,
$D_{min}=$20m,
$v^{fly}_{u}=20$m/s,
$Q=200$,  $K=5$,
$q^{uz}=200$m, $q^{z}_{H}=20$km,
$N^{u}=3$, and $N^{h}=7$.
The remaining parameters refer to \cite{7,10118722,10007714,10713326}.
\begin{figure*}[htbp]
	\centering
	\begin{subfigure}[b]{0.3\textwidth}
		\centering
		\includegraphics[width=1\linewidth]{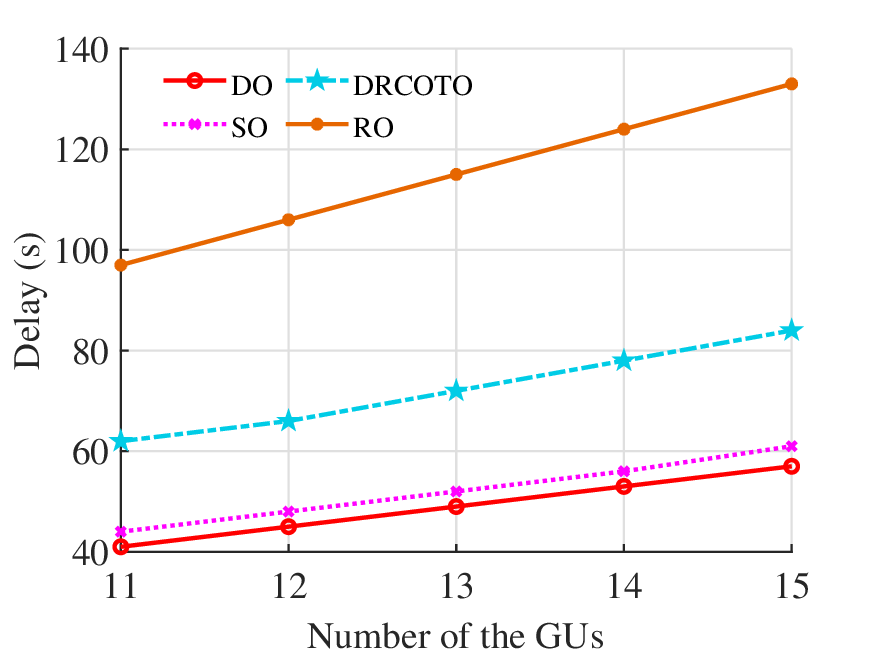}
		\captionsetup{justification=centering, singlelinecheck=true}
		\caption{The optimized delay. \\  \quad}
		\label{subfig:1}
	\end{subfigure}
	\hfil
	\begin{subfigure}[b]{0.3\textwidth}
		\centering
		\includegraphics[width=1\linewidth]{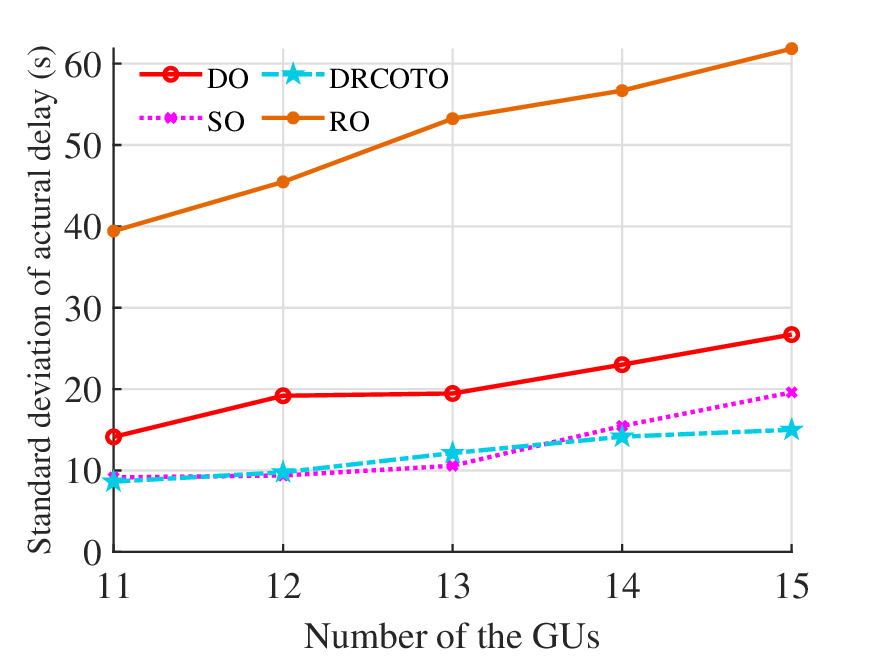}
		\captionsetup{justification=centering, singlelinecheck=true}
		\caption{The standard deviation of actual delays related to optimized delays.}
		\label{subfig:2}
	\end{subfigure}
	\hfil
	\begin{subfigure}[b]{0.3\textwidth}
		\centering
		\includegraphics[width=1\linewidth]{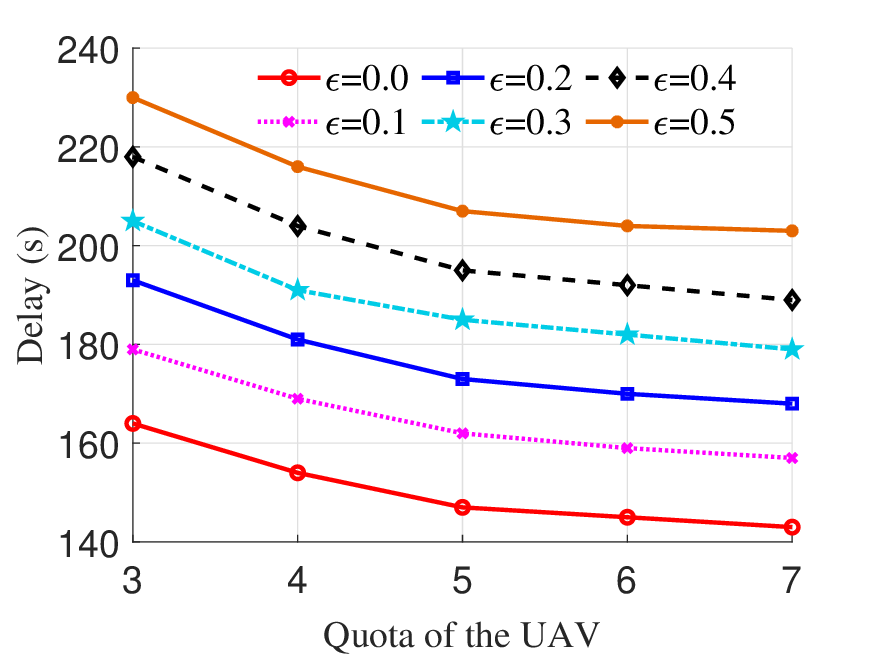}
		\captionsetup{justification=centering, singlelinecheck=true}
		\caption{The relationships between the delay and quota of the UAV.}
		\label{subfig:3}
	\end{subfigure}
	\caption{Comparison of multi-dimensional performance of various optimization methods.}
	\label{fig:1-3}
\end{figure*}

Fig. \ref{fig:12} shows the initial and optimized UAV trajectories in LAN.
The initial and terminal positions of the UAVs are fixed. 
It is observed that all the optimized UAV trajectories curve toward the center of the map. 
This is because the central area of the map has a relatively dense distribution of the GUs and 
the HAP is also positioned there. 
Consequently, the UAVs fly toward the center to reduce the transmission delays.

Fig.\ref{fig:1-3} compares results obtained from the deterministic optimization (DO), 
stochastic optimization (SO), 
DRCOTO, and robust optimization (RO) to 
evaluate the performance 
of the 
proposed 
algorithm. 
In detail, task sizes are treated
as deterministic, ignoring the uncertainty entirely
in the DO.
In the SO, the probability distribution of task
sizes is known. 
In the RO, the  maximum potential task sizes 
is considered.

As shown in Fig. \ref{subfig:1}, 
as the number of GUs increases, 
the delay derived from all optimization methods 
shows an upward trend. 
This is because a larger number of GUs 
leads to more data to be processed, 
which in turn causes the higher delay.
With the same number of GUs, 
the optimized delays of the four methods increase in sequence, 
i.e., DO, SO, DRCOTO, and RO. 
This indicates that the scenarios considered 
by these methods become progressively more conservative. 

Fig. \ref{subfig:2} shows
the standard deviation of actual delays 
related to optimized delays. 
The  "actual delays"
specifically refers to 
the measured delays derived from 
processing five distinct datasets 
with the optimized computation offloading strategy and  UAV trajectories.
As illustrated in Fig. \ref{subfig:2}, 
the standard deviation of actual delays increases 
with the growing number of GUs. 
This is because
the deviation between the actual and optimized task sizes  expands gradually 
with more GUs, 
thereby leading to a higher standard deviation.
Besides, 
with the same number of GUs, 
the standard deviations of the RO, DO, SO, and DRCOTO 
exhibit a sequentially decreasing trend. 
The DRCOTO method yields the smallest standard deviation, 
indicating its superior stability.
It is concluded that 
compared with other benchmarks, 
the DRCOTO  achieves the best balance between 
the delay optimization and robustness.

Fig. \ref{subfig:3} depicts 
the relationships between the delay and quota of the UAV.
As shown in Fig. \ref{subfig:3}, 
the delay decreases as the quota of UAVs increases. 
This is attributed to the fact that 
the computation delay of UAVs is lower than that of GUs. 
With larger quotas of UAVs,
more tasks can be computed on UAVs, 
thereby reducing the delay.
Furthermore, the optimized delay increases with 
the expanding radius of the uncertainty set.
This is because a larger radius covers more adverse possible probability distributions, 
particularly those with larger task sizes, 
which ultimately leads to an increment in the delay.

\section{CONCLUSIONS}\label{CONCLUSIONS}
\vspace{2mm}
In this paper, 
we highlight the cooperation mechanisms of 
UAVs and HAPs in the LAN to provide offloading services for GUs.
The uncertainty sets are established based on the $L_{1}$ norm metric 
to measure uncertainties in task sizes. 
Based on the constructed uncertainty sets,
a DRO problem is formulated to 
minimize the worst case delay 
by jointly optimizing offloading strategies 
and UAV trajectories.
Subsequently, the DRCOTO algorithm is developed 
based on the BD and SCA to solve the proposed DRO problem. 
Simulation results demonstrate
the DRCOTO algorithm outperforms the DO, SO, and RO methods 
in achieving a trade-off between the optimized delay and robustness.
\vspace{2mm}
\bibliographystyle{IEEEtran}
\bibliography{REFERENCES.bib}
\end{document}